\documentclass[a4paper,12pt]{article}
\pdfoutput=1

\usepackage{jheppub}
\usepackage{bm,latexsym,amsmath,amssymb,amsfonts,mathtools,mathrsfs}

\usepackage{graphicx}
\usepackage{wrapfig}
\usepackage{hyperref}
\usepackage[hang]{subfigure}
\usepackage{color}
\usepackage{multirow}

\usepackage{todonotes}
\definecolor{palatinate}{RGB}{128, 49, 123}

\newcommand{\be}{\begin{equation}}
\newcommand{\ee}{\end{equation}}
\newcommand{\beal}{\begin{aligned}}
\newcommand{\eeal}{\end{aligned}}
\newcommand\bea{\begin{eqnarray}}
\newcommand\eea{\end{eqnarray}}
\newcommand{\bec}{\begin{cases}}
\newcommand{\eec}{\end{cases}}

\title{Seeded vacuum decay with Gauss-Bonnet}

\author[a,b]{Ruth Gregory}
\author[a,1]{Shi-Qian Hu}
\note{Corresponding author}

\emailAdd{ruth.gregory@kcl.ac.uk}
\emailAdd{shiqian.hu@kcl.ac.uk}

\affiliation[a]{Department of Physics, King's College London,
The Strand, London WC2R 2LS, UK}
\affiliation[b]{Perimeter Institute, 31 Caroline Street North, Waterloo, 
ON, N2L 2Y5, Canada}

\date{\today}

\abstract{ We investigate false vacuum decay catalysed by black holes under the influence of the higher order Gauss-Bonnet term.
We study both bubble nucleation and Hawking-Moss types of phase transition in arbitrary dimension. 
The equations of motion of ``bounce'' solutions in which bubbles nucleate around arbitrary dimensional black holes are found in the thin wall approximation, and the instanton action is computed. The headline result that the tunnelling action for static instantons is the difference in entropy of the seed and remnant black holes is shown to hold for arbitrary dimension. We also study the Hawking-Moss transition and find a picture similar to the Einstein case, with one curious five-dimensional exception (due to a mass gap). In four dimensions, we find as expected that the Gauss-Bonnet term only impacts topology changing transitions, i.e.\ when vacuum decay removes the seed black hole altogether, or in a (Hawking-Moss) transition where a black hole is created. In the former case, topology changing transitions are suppressed (for positive GB coupling $\alpha$), whereas the latter case results in an enhanced transition.}

\begin{document}

\maketitle
\section{Introduction}

It is well known that while classically a particle in a local minimum of potential
is stable to small perturbations, the same is not necessarily true quantum 
mechanically -- it depends on whether the particle sits in a global or only a local minimum.
If the latter, a process of quantum tunnelling occurs, where the particle ``borrows''
energy by quantum uncertainty to emerge on the other side of the barrier. While
well-tested in non-relativistic systems, tunnelling from such a local minimum or
{\it false vacuum} in Quantum Field Theory lacks experimental verification. 

By far the most intuitive picture of vacuum decay is that of a first order
phase transition: a bubble of true vacuum fluctuates into existence inside
a false vacuum state, then expands to convert to true vacuum everywhere.
This picture, best described in a sequence of papers by Coleman and collaborators 
\cite{Coleman:1977py,Callan:1977pt} (see also \cite{Kobzarev:1974cp}) originally focused 
on the field theory aspect; considering the problem more holistically however
mandated including gravity due to the impact of the vacuum energy.
As described in \cite{Coleman:1980aw}, this requires using the semi-classical 
partition function approach developed by 
Gibbons and Hawking \cite{Gibbons:1979xm}
where the path integral
is assumed to be dominated by its saddle points, or, solutions to the Euclidean 
Einstein/field theory equations; these are found, and the action of the bubble is
computed with the amplitude for decay being, to leading order,
\be
{\cal P} \sim e^{-{\cal B}} = e^{-(I-I_0)} \;,
\ee  
where $I$ denotes the Euclidean action of the bubble instanton solution, and $I_0$ 
the action of the false vacuum background. 

Coleman and de Luccia (CDL) originally computed the gravitational action of a bubble of
the maximally symmetric spacetime associated with the vacuum energy -- de Sitter for 
positive vacuum energy, Minkowski for zero vacuum energy, and anti-de Sitter 
for negative vacuum energy. This, however, is a very idealised picture of the vacuum,
with no structure or inhomogeneity. In \cite{Gregory:2013hja,Burda:2015yfa}, 
an extension of the CDL approach was
considered, where a bubble was assumed simply to be spherical, i.e.\ with SO(3) 
symmetry. A generalisation of the Birkhoff theorem in the presence of branes (the
bubble wall) \cite{Bowcock:2000cq} shows that the general solution of a bubble with SO(3) symmetry
is Schwarzschild (A/dS) -- i.e.\ the general bubble will also surround a black hole,
and in principle can have different `masses' inside and outside the bubble\footnote{Note
that the mass here refers to the local tidal forces, or Riemann curvature, which have
the form $GM/r^3$ locally.}. 
The finding of \cite{Gregory:2013hja,Burda:2015yfa}
(see also \cite{Burda:2015isa,Burda:2016mou,Cuspinera:2018woe,Tetradis:2016vqb,Mukaida:2017bgd,Dai:2019eei,Oshita:2018ptr,DeLuca:2022cus} for
applications to Higgs decay) was that the black hole could significantly reduce the
instanton action, resulting in a strongly enhanced probability of false vacuum decay
in the presence of primordial black holes \cite{Hawking:1971ei,Carr:1974nx,Khlopov:2008qy,Tetradis:2016vqb,Canko:2017ebb,Dai:2019eei}. 
The bounce formalism not only gives the transition rate but also the initial conditions for the real-time evolution of the bubble, which is obtained via analytic continuation of the Euclidean solution. 
A discussion of the correspondence between the Euclidean and Lorentzian sections can be found e.g.\ in \cite{Coleman:1980aw,Brown:2007sd,Burda:2016mou}.

\begin{figure}[ht]
\centering
\includegraphics[width=10cm]{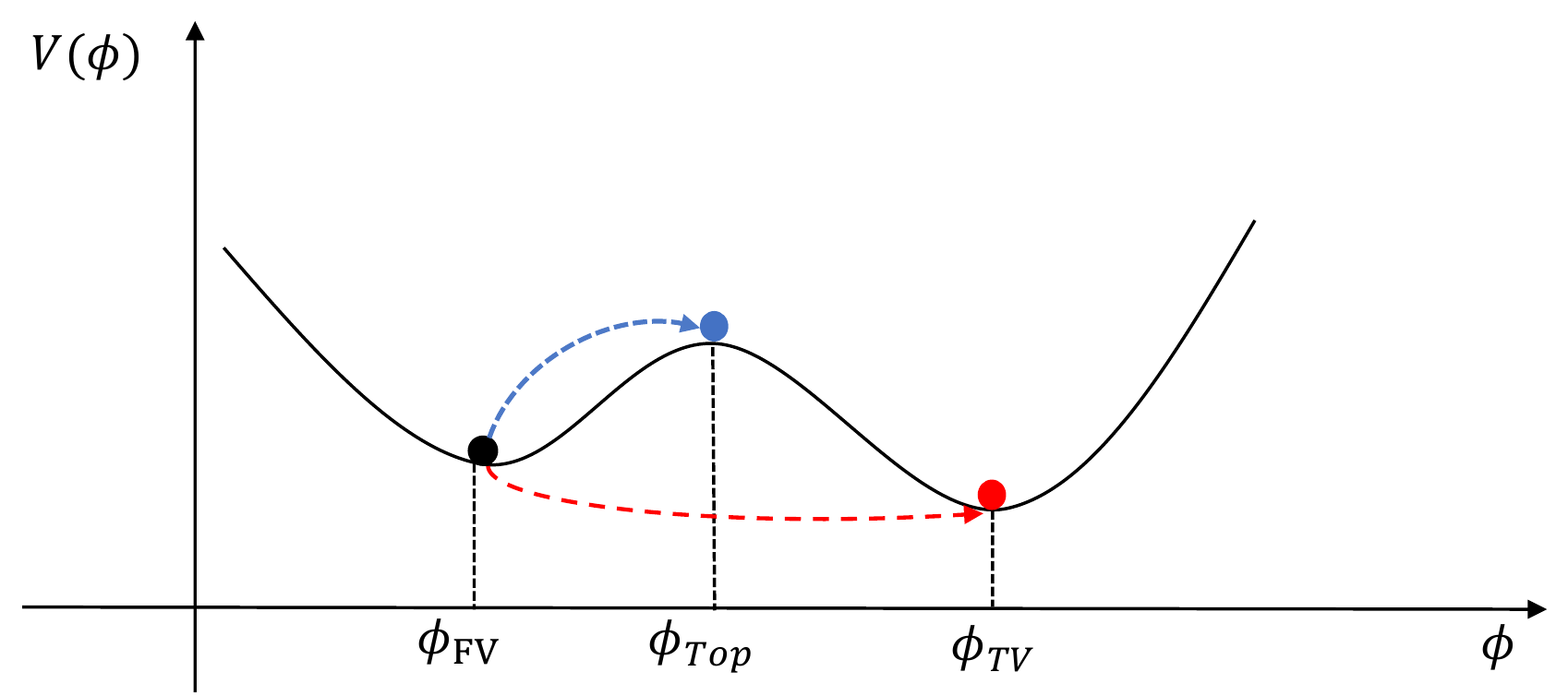} 
\caption{The schematic plot of a one-dimensional potential that has two different minima, with the higher (lower) representing false (true) vacuum, labelled as $\phi_{FV}$ and $\phi_{TV}$, respectively. The top of the potential barrier is labelled as $\phi_{Top}$. A particle initially at the local minimum can arrive at the global minimum by two mechanisms: (\romannumeral1) Bubble nucleation (red) - quantum tunnelling to true vacuum or (\romannumeral2) HM transition (blue) - climbing to the top of the potential and then rolling down to the true vacuum.  }
\label{fig: potential}
\end{figure}

Bubble nucleation is not the only decay channel for a false vacuum. The Hawking-Moss (HM) instanton describes an up-tunnelling of the false vacuum to a local maximum of the 
potential, allowing a roll-down to the true minimum \cite{Hawking:1981fz}. 
The two tunnelling processes can be best illustrated via the cartoon of a particle in a one-dimensional potential barrier, as shown in Fig.\ \ref{fig: potential}.
The first order, or bubble nucleation, process corresponds to a particle transitioning from the local minimum on one side of the barrier to the other side. 
In the HM transition on the other hand, the particle ``jumps'' to the top of the potential, then rolls down to the true minimum. This HM channel was shown to be relevant
for potentials with a gently varying barrier (the bound being on the second derivative
of $V(\phi)$ at the local maximum \cite{Hawking:1981fz}). 
The impact of black holes on the HM transition was considered in \cite{Gregory:2020cvy,Gregory:2020hia}, 
see also \cite{Gomberoff:2003zh}, with an outcome broadly similar to the bubble nucleation process, namely, 
that black holes enhance the probability of tunnelling. There is one crucial difference however that is 
relevant for the study here: For bubble nucleation, there can only be a remnant black hole if there is a 
black hole seed to nucleate the decay. For HM transitions however, it is possible to jump from a pure 
de Sitter false vacuum to a Schwarzschild de Sitter configuration at the top of the potential barrier. 

It is worth noting that the classic results of Coleman and de Luccia are obtained within the 
\emph{thin wall} approximation, in which the vacuum transitions completely and instantaneously 
across the wall. The thin wall approximation, while excellent for testing gravitational 
effects, is only a good approximation for vacuum transitions in scalar fields when the potential barrier is sharp. 
For vacuum energy sourced by a scalar field, 
there is a region around the ``wall'' where the 
scalar rolls to its vacuum value and for the important physical application of Higgs vacuum decay, the ``wall'' is a rather broad scalar distribution around the black hole. The early results of 
\cite{Burda:2015yfa,Burda:2015isa} stayed strictly within the semi-classical approximation, and did not
include possible thermal corrections to the Higgs potential due to the black hole temperature. 
Later work 
\cite{Hayashi:2020ocn,Strumia:2022jil,Miyachi:2021bwd,Shkerin:2021zbf,Briaud:2022few,Shkerin:2021rhy} has explored
these corrections, however there is still a lack of
consensus as to the precise magnitude of any suppression. 
The problems lie around the choice of vacuum for the Higgs, as well as the validity of calculational methods.
In \cite{Hayashi:2020ocn}, Hayashi et al.\ perform a Euclidean computation of the tunnelling action including the local thermal corrections to the Higgs potential and their backreaction, concluding that while there is some lowering of the instanton decay probability, it is not suppressed completely. 
While this is an explicit calculation, one can critique the Euclidean approach as in \cite{Strumia:2022jil}, where Strumia uses a sphaleron approximation for the instanton to derive expressions capturing the parametric dependancy in certain limits, resulting in an apparent suppression of the process altogether. While phenomenologically compelling, \cite{Strumia:2022jil} focusses on the primary parametric dependence in key limits, and does not address important prefactors in these expressions, which can be key in the comparison of evaporation to tunnelling. In \cite{Shkerin:2021zbf}, Shkerin and Sibiryakov perform a first principles calculation for all of the quantum vacua (Boulware, Hartle-Hawking and Unruh) which indicates that tunnelling is still significant, however the argument is performed in two spacetime dimensions, although the authors argue that the result can be extrapolated to four dimensions. However, this would require care in taking boundary conditions on the tunnelling solutions corresponding to different initial vacua. 
In essence, the problem centres around how to correctly incorporate gravity in non-perturbative processes that do not even have experimental verification in the absence of gravity. While these concerns and arguments are important and merit continuing investigation, in this work we are interested in the effect of higher curvature gravitational terms on instantons in general, thus will focus on exact solutions which transition between different \emph{cosmological constants}, and thus will not be considering a direct application to Higgs vacuum decay. Among the theories containing higher-order corrections the most natural are those that maintain second order equations of motion -- the Lovelock terms \cite{Lovelock:1971yv} -- and we focus here on the Gauss-Bonnet (GB) term \cite{Lanczos:1938sf,Zumino:1985dp,Zwiebach:1985uq}
\be
{\cal L}_{GB} = R^2 - 4 R_{ab}^2 + R_{abcd}^2
\label{GBterm}
\ee
that is known in 4D not to alter the dynamics of
the theory. It may seem strange to consider a correction that does not alter the dynamics
of a system, however, when computing the probability of decay, it is the \emph{action}
of a process that is relevant, and the GB term will alter the action of a solution, 
potentially affecting the tunnelling amplitude. (Note, we 
are not considering the Glavan-Lin approach \cite{Glavan:2019inb}, 
as this has several problems
\cite{Arrechea:2020gjw,Gurses:2020rxb, Ai:2020peo}.)

Our main finding is that due to the topological nature of the GB term,
the results for black hole seeded decay are not, in the main, compromised by the 
addition of this higher derivative term in 4D. However, we find that the presence of the
GB term almost always suppresses topology-changing transitions. The main exception to this
result is HM decay from pure de Sitter space, where the GB term can 
provide an enhancement to the decay for small black holes.

The paper is organized as follows: In Sec.\ \ref{sec:bubble nucleated decay} we set up the 
formalism of instanton and compute the instanton action in order to show the effects of GB term. 
In Sec.\ \ref{sec:HM instanton}, we discuss the HM transitions and display applications in higher 
dimensions ($d\geq 5$), and explain how GB terms contribute when topology changes. 
In Sec.\ \ref{sec:4d}, we restrict to $D=4$ where bubble solutions are identical to Einstein 
gravity and examine the difference with that in Einstein's. Finally, we conclude our results 
in Sec.\ \ref{sec:discussion}. 
We work throughout in the \emph{thin wall} approach of Coleman
and de Luccia \cite{Coleman:1980aw}, as this captures the gravitational aspects of tunnelling in a 
straightforward analytic fashion, thus should isolate effectively the impact of the GB term.
We use natural units ($c=\hbar=1$) throughout the paper.

\section{Bubble-nucleated decay}\label{sec:bubble nucleated decay}

We first briefly review the procedure for finding the instanton, noting how
the GB term impacts the argument in general and deriving the action in the 
presence of the GB term.
Gauss-Bonnet instantons with $O(D)-$symmetry were explored in \cite{Charmousis:2008ce} in the context of stability of the different branches of the Einstein-Gauss-Bonnet (EGB) vacua, here we are looking for solutions with a black hole, hence $O(D-1)$ symmetry.

Recall that EGB gravity is the simplest of the Lovelock extensions to General Relativity
\cite{Lovelock:1971yv}, with an action given by
\be
I = - \frac1{16\pi G} \int_{\cal M} d^D x \sqrt{g} 
\left [ {\cal R}- 2 \Lambda +\alpha {\cal L}_{GB}\right] 
\label{EGBaction}
\ee
where $\alpha$ is the (dimensionful) coupling constant of the GB term, 
${\cal L}_{GB}$, defined in \eqref{GBterm}.
The general (Euclidean) spherically symmetric vacuum solution is derived in \cite{Boulware:1985wk},
\be \label{lineelement}
ds^2= f(r)d\tau^2+f^{-1}(r)dr^2+r^2d\Omega^{2}_{D-2}
\ee
where $d\Omega^{2}_{D-2}$ gives the metric of the unit $S^{D-2}$ and we take the
`Einstein' branch solution \cite{Wheeler:1985nh,Wheeler:1985qd,Myers:1988ze}:
\be\label{metricfunc}
f(r)=1+\frac{r^2}{2\tilde{\alpha}}\left(1-\sqrt{1+
\frac{8\tilde{\alpha} \Lambda}{(D-1)(D-2)}+\frac{4\tilde{\alpha}\mu}{r^{D-1}}}\right)
\ee
with $\tilde{\alpha}=(D-3)(D-4)\alpha$, and $\mu$ related to the mass through
the standard Myers-Perry formula \cite{Myers:1986un}:
\be
\mu = \frac{16\pi G M}{(D-2) {\cal A}_{D-2}}
\ee
with ${\cal A}_{D-2}$ being the area of a unit $(D-2)-$sphere.
Note that as $\tilde{\alpha}\to 0$, we recover
the Schwarzschild solution. Note also that if $\alpha<0$, there are r\'egimes of parameter space where the argument of the root in \eqref{metricfunc} becomes negative, hence the solution is unphysical. While what follows is independent of the sign of $\alpha$, we will focus on $\alpha>0$ for discussion.

A bubble separating false vacuum from true vacuum is expected to have SO(3) 
symmetry, and a generalisation of the Birkhoff theorem in the presence of branes
\cite{Bowcock:2000cq} shows that the general bulk solution is Schwarzschild (A/dS) with the bubble 
surrounding the black hole. This theorem was also demonstrated for GB gravity
in \cite{Wiltshire:1985us,Charmousis:2002rc,2010AIPC.1241..521B}, with the general brane junction conditions derived in
\cite{Davis:2002gn}. Although the thin wall approximation is an idealised description
of bubble nucleated vacuum decay, it will suffice for the exploration of the 
impact of the GB term.

Following the steps of \cite{Gregory:2013hja}, we seek a Euclidean bubble solution $R(\lambda)$ 
separating two black hole spacetimes with exterior mass $M_+$ (the seed) and
false vacuum energy $\Lambda_+$, and an interior mass $M_-$,
and true vacuum energy $\Lambda_-$.
The regions exterior and interior to the wall are given by
\begin{equation}\label{Emetric}
ds^{2}_\pm=f_\pm d\tau^{2}_{\pm}+f_\pm^{-1}dr^{2}+r^{2}d{\Omega}^2_{D-2},
\end{equation} 
where 
\be\label{metricfunctions}
f_\pm =1+\frac{r^2}{2\tilde{\alpha}}\left(1-\sqrt{1+
\frac{8\tilde{\alpha} \Lambda_\pm}{(D-1)(D-2)}
+\frac{4\tilde{\alpha}\mu_\pm}{r^{D-1}}}\right),
\ee
with the boundary $R(\lambda)$ being given by the generalisation of the Israel prescription 
to EGB gravity as we now describe.

The full action of the composite system is $I = I_{bulk} + I_{brane}$, where
\be
\beal
I_{bulk} = - \frac1{16\pi G} \Biggl ( &\int_{\cal M_+} d^D x \sqrt{g_+} 
\left [ {\cal R}_+- 2\Lambda_+ +\alpha {\cal L}^+_{GB}\right] \\
&+\int_{\cal M_-} d^D x \sqrt{g_-} 
\left [ {\cal R}_- - 2\Lambda_- +\alpha {\cal L}^-_{GB}\right] \Biggr )
\eeal
\label{Ibulk}
\ee
represents the contribution from the spacetime away from the wall, and
\be
\beal
I_{brane}=
\int_{\mathcal{W}} d^{D-1} x \sqrt{h} \left (  \sigma +
\frac{1}{8 \pi G} \left( \Delta K -2\alpha \left[2\mathcal{G}_{ab} \Delta K^{ab}
-\Delta \mathcal{J}\right] \right) \right)
\end{aligned}
\label{Ibrane}
\end{equation}
represents the net contribution from the wall, including both the wall tension 
$\sigma$, and the geometrical Gibbons-Hawking terms from the boundary
submanifolds on each side: $\partial {\cal M}_+ = \partial {\cal M}_- = {\cal W}$.
Here, $K_{_\pm ab}$ is the extrinsic curvature of the wall as determined from 
the local embedding into ${\cal M}_\pm$, ${\cal G}_{ab}$ is the local intrinsic
Einstein tensor of the wall, and ${\cal J}$ is a cubic extrinsic tensor (see \cite{Davis:2002gn,Brihaye:2008xu,Brihaye:2008ns,Liu:2008zf})
:
\be
{\cal J}_{ab} = \frac13 \left ( 2 K K_{ac} K^c_{\; b} + K_{cd}K^{cd} K_{ab} 
-2K_{ac}K^{cd}K_{db} - K^2 K_{ab} \right).
\ee

\subsection{The instanton solution}

In order to find the instanton, we have to solve the equations of motion for
the bubble wall, found by varying \eqref{Ibulk} and \eqref{Ibrane}.
As described in \cite{Davis:2002gn}, the generalised Israel equations are:
\be
\Delta K_{ab} - \Delta K h_{ab} + 2 \alpha \left [3 \Delta{\cal J}_{ab}
- \Delta{\cal J} h_{ab} - 2 {\cal P}_{acbd} \Delta K^{cd} \right] = 8 \pi G \sigma h_{ab}
\ee
with 
\be
{\cal P}_{abcd} = {\cal R}_{abcd} + 2  {\cal R}_{b[c} h_{d]a} 
-2  {\cal R}_{a[c} h_{d]b} +  {\cal R} h_{a[c} h_{d]b}
\ee
the divergence free part of the intrinsic Riemann tensor.

Because of the SO(3) symmetry of the bubble, these rather complex relations
simplify a little; we write the bubble wall as $R(\lambda)$, where $\lambda$ is the proper 
time of an observer comoving with the wall:
\begin{equation}\label{EOM}
f _{\pm}(R)\dot{\tau}^{2}_{\pm}+\frac{\dot{R}^2}{f_{\pm}(R)}=1,
\end{equation}
and choosing a normal always pointing towards increasing $R$ 
for $\dot{\tau}_\pm>0$,
\begin{equation}
n_{\pm}=\left(-\dot{R} d \tau_{\pm}+\dot{\tau}_{\pm} d r_{\pm}\right),
\end{equation}
gives, after some algebra, the junction condition as
\begin{equation}
\left ( f_+\dot{\tau}_+ -f_-\dot{\tau}_-\right) 
\left [ 1 + \frac{2\tilde{\alpha}}{R^2} - \frac{4\tilde{\alpha}}{3} \frac{{\dot R}^2}{R^2} \right]
- \frac{2\tilde{\alpha}}{3R^2} \left ( f_+^2\dot{\tau}_+ -f_-^2\dot{\tau}_-\right) 
=- \frac{8\pi G \sigma R}{(D-2)},
\end{equation}
where $\sigma$ is the surface tension of the wall. Substituting 
$f_\pm \dot{\tau}_\pm = \sqrt{f_\pm - \dot{R}^2}$ from \eqref{EOM} leads to a 
rather lengthy cubic Friedmann-like equation for $X = (1-\dot{R}^2)/R^2$,
\be
\beal
0 &=  \tilde{\alpha}^2 X^3 +  \tilde{\alpha} X^2 \left ( \frac32
- \frac{(S_+^2-S_-^2)^2}{256 \tilde{\alpha} \bar{\sigma}^2} \right)\\
&+ \frac{X}{4} \left ( 3 - \frac{3(S_+^2+S_-^2)}{8}
- \frac{(S_+^2 - S_-^2)(3S_+^2-3S_-^2 + S_+^3 - S_-^3)}
{192 \tilde{\alpha} \bar{\sigma}^2} \right ) \\
&+ \frac{(8 - 3S_+^2-3S_-^2 - S_+^3 - S_-^3)}{64 \tilde{\alpha}} 
- \frac{(3S_+^2-3S_-^2 + S_+^3 - S_-^3)^2}{9216 \tilde{\alpha}^2 \bar{\sigma}^2}
-\frac{9\bar{\sigma}^2}{16}
\eeal
\label{cubicFriedmann}
\ee
where $\bar{\sigma} = 2\pi G\sigma /(D-2)$ economises on notation, and 
\be
S_\pm = \sqrt{1+\frac{8\tilde{\alpha} \Lambda_\pm}{(D-1)(D-2)}
+\frac{4\tilde{\alpha}\mu_\pm}{r^{D-1}}}
= \sqrt{ 1 + \frac{4\tilde{\alpha}}{r^2} (1-f_{_\pm E})}
\ee
represents the geometry on each side of the wall and is related to the standard Einstein
potential $f_E = 1 - \mu/r^{D-3} -2 \Lambda r^2/(D-1)(D-2)$. 
Note that $S\simeq 1 + {\cal O}(\tilde{\alpha})$, so as $\tilde \alpha \to 0$,
\eqref{cubicFriedmann} reduces to
\be
\beal
&\frac{9}{16} \left [ X -\bar{\sigma}^2 - \frac{(1- \bar{f}_E)}{R^2} 
- \frac{(\Delta f_E)^2}{16 \bar{\sigma}^2R^4} \right ] = 0\\
&\Rightarrow \qquad \left(\frac{\dot{R}}{R}\right)^{2}
= \frac{\bar{f}_E}{R^2} -\bar{\sigma}^2 
- \frac{(\Delta f_E)^2}{16 \bar{\sigma}^2R^4}
\eeal
\ee
as required. (Here, $\bar{f}_E\equiv(f_{E-}+f_{E+})/2$ and $\Delta f \equiv f_{E+}-f_{E-}$.) 
Since $\tilde{\alpha} = 0$ for $D=4$, we see that the EGB equations
of motion for the wall indeed reduce to the Einstein equations, as expected for a
physical system. Therefore, in 4D, the motion of the wall is as the Einstein case
described in \cite{Gregory:2013hja,Burda:2015yfa}: for a given seed mass $M_+$, there are a range of bubble 
solutions with remnant masses $M_-$, however, we expect a unique lowest action 
solution with specific mass $M_-$ that we identify as the instanton with a remnant black
hole of mass $M_-$.

\subsection{Computing the instanton action}

In order to identify the lowest action solution (bearing in mind that even in 4D, the 
GB term could contribute to the action), we must now compute the bulk
and wall contributions to the action, then subtract the background action of the initial state.

\subsubsection{The background seed action}

We start by computing the action for the background solution of mass $M_+$,
and bare vacuum energy $\propto \Lambda_+$; this is simply a matter of computing the
bulk term \eqref{Ibulk}, bearing in mind two key points. The first is that if we are in 
vacuum or AdS spacetime ($\Lambda_+ \leq 0$) then we must truncate our bulk integral
at some finite radius that we choose to label $r_c$. For a positive cosmological
constant, $r_c$ will be the upper limit of integration at the cosmological horizon.
The second point is that we must perform the integration at \emph{arbitrary}
Euclidean time periodicity $\beta$, hence at an event horizon, there will, in general,
be a conical deficit\footnote{Although the presence of a conical deficit can be disconcerting, the singularity is mild, and can be resolved by a process analogous to the discussion of the Dirac delta-distribution (as described in the appendix of \cite{Gregory:2013hja}). As discussed in \cite{Gibbons:1979xm}, the horizon is a fixed point of the isometry of the metric – a “bolt” – which has codimension 2, hence is not a boundary of the spacetime in the strict sense.}, thus we must generalise the argument of \cite{Gregory:2013hja}
to include the GB term. 

Computing the integrand of \eqref{EGBaction} for the metric \eqref{lineelement}
we see that, on-shell, the bulk integrand is a total derivative:
\be
\sqrt{g} \left [ {\cal R} - 2\Lambda +\alpha {\cal L}_{GB}\right]
= - \left \{ r^{D-4} f' \left [ r^2 + \frac{2(D-2) \tilde{\alpha}} {(D-4)} (1-f) \right] \right\} '.
\label{onshellbulk}
\ee
Giving a contribution of $I_c-I_h$, where
\be
I_c = \frac{\beta {\cal A}_{D-2}}{16\pi G} r_c^{D-4} f'(r_c) 
\left ( r_c^2 + \frac{2(D-2) \tilde{\alpha}} {(D-4)} (1-f(r_c)) \right) 
\label{rccontribution}
\ee
at the upper $r_c$ limit, and at the black hole horizon
\be
I_h = \frac{\beta {\cal A}_{D-2}}{16\pi G} \times
\frac{4\pi}{\beta_h} r_h^{D-2}
\left ( 1 + \frac{2(D-2) \tilde{\alpha}}{(D-4)r_h^2} \right)
= \frac{\beta}{\beta_h} S_{BH}
\label{rhcontribution}
\ee
the contribution is proportional to the entropy. Note we have replaced the 
derivative of $f$ at the horizon with $f'(r_h) = 4\pi/\beta_h$, an expression 
involving the critical horizon periodicity $\beta_h$ for which there is no conical singularity at the horizon. 

We now complete the computation by taking into account possible conical
defects at a horizon. Following the approach of \cite{Gregory:2013hja} (see also \cite{Li:2023men}), we note that a conical
deficit metric has the local form $d\rho^2 + \lambda^2\rho^2 d\theta^2$, where 
$\lambda\neq1$ and the periodicity of $\theta$ is $2\pi$. The deficit angle for 
this metric is $\delta = 2\pi (1-\lambda)$. We write the black hole metric locally
at the horizon as
\be
ds^2 = A^2(\rho) d\tau^2 + d\rho^2 + C^2(\rho) d\Omega_{D-2}^2
\label{conicalform}
\ee
where $\rho^2 = 4( r-r_h ) / f'(r_h)$, and $A'(0) = 2\pi / \beta_h$.
Comparing this with the canonical form above ($\theta = 2\pi \tau/\beta$),
we see that $\lambda = \beta/\beta_h$, hence the conical deficit is
\be
\delta = 2\pi \frac{\beta_h-\beta}{\beta_h}.
\ee

To compute the contribution of the conical deficit to the action, we split our
bulk integral into an integral from a proper distance $\varepsilon$ from
the horizon, and smooth out the cone inside $\varepsilon$, by having $A$
smoothly interpolate between being regular at the origin, $A'(0) = 2\pi/\beta$,
and having the black hole geometry value $A(\varepsilon) = 2\pi/\beta_h$
at $\rho=\varepsilon$. We then take the limit $\varepsilon\to0$ to obtain the
action with the conical singularity.

Computing the terms in the EGB Lagrangian for the metric 
\eqref{conicalform}:
\be
\beal
{\cal R}  &=  -2 \frac{A''}{A} -2 (D-2) \left ( \frac{A'C'}{AC} -\frac{C''}{C}\right)
+ (D-2)(D-3) \frac{(1-C^{\prime2})}{C^2} \\
\alpha {\cal L}_{GB} &= -4 (D-2)\tilde{\alpha} \Biggl [  
\frac{(1-C^{\prime2})}{(D-4)C^2} \frac{A''}{A}
+ \frac{(1-C^{\prime2})}{C^2} \left ( \frac{A'C'}{AC} + \frac{C''}{C}\right)\\
&\hskip 3cm
- (D-5) \frac{(1-C^{\prime2})^2}{4C^4} \Biggr]
\eeal\ee
and using $C = r_h + {\cal O}(\rho^2)$, shows that near the horizon
\be 
AC^{D-2} \left [ {\cal R} - 2\Lambda + \alpha{\cal L}_{GB} \right] \sim
-2 \left (A' \left [C^{D-2} + \frac{2\tilde{\alpha}(D-2) } {(D-4)} C^{D-4}\right ] \right) '+ {\cal O}(\rho)
\ee
hence
\bea
\int_0^\varepsilon d\rho AC^{D-2} \left [ {\cal R} - 2\Lambda + \alpha{\cal L}_{GB} \right] &=&
-2 r_h^{D-2} \left ( 1 + \frac{2(D-2)\tilde{\alpha}}{(D-4)r_h^2} \right) 
\left [ A'(\varepsilon)-A'(0) \right] + {\cal O}(\varepsilon^2) \nonumber \\
&=& -2 \left ( \frac{4G S_{BH}}{{\cal A}_{D-2}} \right) 
\frac{2\pi(\beta - \beta_h)}{\beta \beta_h}
\eea
Thus the full contribution to the bulk action including the conical deficit is
\be
\beal
-\int d^D x \sqrt{g}\frac{{\cal L}_{EGB}}{16\pi G} &= 
-\lim_{\varepsilon\to0}\frac{\beta{\cal A}_{D-2}}{16\pi G} \left [ 
\int_0^\varepsilon AC^{D-2} {\cal L}_{EGB} d\rho
+ \int^{r_c}_{r_h+{\cal O}(\varepsilon^2)} \hskip -1cm r^{D-2} {\cal L}_{EGB} dr \right]\\
&= \frac{(\beta - \beta_h)}{\beta_h} S_{BH} - \frac{\beta}{\beta_h} S_{BH} + I_c 
= I_c -S_{BH}
\eeal
\ee
The large $r$ contribution from $r_c$ is treated differently depending on whether 
$\Lambda>0$ or not. For $\Lambda>0$, we have a cosmological horizon, and must
therefore take into account any conical deficit as described above. This procedure follows through
in much the same way as for the black hole horizon, and (noting that $I_c<0$ for the cosmological horizon) for a de Sitter seed we obtain
\be\label{diffinaction}
I_{seed} = - S_{CH} - S_{BH}
\ee
thus, just as for Einstein gravity, the action of a black hole in de Sitter space for 
EGB gravity is the sum of the entropies of the horizons. 

If however we are in asymptotically flat or AdS spacetime ($\Lambda<0$ or $\Lambda=0$), we place a cut-off
boundary at $r_c$, and compute the action at that cut-off. Strictly, this means we have 
to also compute a Gibbons-Hawking boundary term for the artificial boundary $r_c$, as well as perform a 
background subtraction to obtain a finite answer. However, in computing the action
of the instanton, we calculate the action of the bubble geometry and subtract the action of the seed geometry;
noting that the bubble and seed geometries are identical outside the bubble wall,
both the contribution from the bulk integral $I_c$ (eq.\ \eqref{rccontribution}) as well 
as the Gibbons-Hawking term and any background subtractions will be the same.
Thus we do not need to explicitly compute these large $r$
terms 
and simply note that both seed and bubble geometries have an identical contribution
at the large $r$ cut-off, which we label $I_C$:
\be
I_{seed} = I_{C} - S_{BH}(M_+,\Lambda_+)
\ee

\subsubsection{The bubble geometry}

For the bubble geometry, the main difference from the previous subsection 
is that we have the bubble wall at $R(\lambda)$. If $R$ varies with $\lambda$,
then the periodicity of the solution is set by the periodicity of $R$. The previous
subsection demonstrates how to compute the bulk action, which is now composed
of two parts, the interior and exterior of the bubble. Using these results, we see
that the bulk contribution to the bubble action, \eqref{Ibulk}, reduces to the boundary
terms at $r_h$ and $r_c$, together with a contribution evaluated at the wall:
\be
\beal
I_{bulk, {\cal W}} = 
& - \frac{{\cal A}_{D-2}}{16\pi G} \int d\lambda R^{D-4} \Bigl [ f'_+ \dot{\tau}_+
\left ( R^2 + \frac{2(D-2)\tilde{\alpha} } {(D-4)} (1-f_+) \right) \\
& -f'_- \dot{\tau}_- \left ( R^2 +  (1-f_-) \right ) \Bigr].
\eeal
\ee
Turning to the wall integral \eqref{Ibrane}, the trace of the Israel equations
\be
(D-1) 8\pi G \sigma = - (D-2)  \Delta K + \frac{2\tilde{\alpha}} {(D-3)} \left [
{\cal G}_{ab} \Delta K^{ab} - \Delta {\cal J} \right]
\ee
gives the wall integral as
\be
I_{\cal W}= \frac{{\cal A}_{D-2}}{8\pi G} \int d\lambda \frac{R^{D-2}}{(D-1)}
\left ( \Delta K -\frac{6\tilde{\alpha}}{(D-3)(D-4)} \left[2\mathcal{G}_{ab} \Delta K^{ab}
-\Delta \mathcal{J}\right] \right) .
\ee
It then proves useful to define
\be
K_0 = K_{\lambda\lambda} = \frac{f' - 2\ddot{R}}{2f\dot{\tau}} \; ,\qquad
K_1  = \frac{f\dot{\tau}}{R} \; , \qquad
p_0 = \frac{(1-\dot{R}^2)}{2R^2} \; , \qquad
p_1 = \frac{\ddot{R}}{R}
\label{kays}
\ee
here, $K_1$ is the angular part of the extrinsic curvature, and $p_0$, $p_1$ appear in 
the intrinsic Einstein tensor. The Israel conditions then give the following relation
between the timelike and spacelike extrinsic curvature differentials:
\be
\Delta K_1 = \Delta K_0 + 2 \tilde{\alpha}
\left [ 2p_0  \Delta K_0 - 2p_1  \Delta K_1- \Delta (K_0K_1^2) 
+  \Delta K_1^3 - 6p_0 \Delta K_0 \right].
\ee
Using these relations, after some algebra we find the wall integral can be reduced to

\be
\resizebox{0.90\hsize}{!}{$\beal
I_{\cal W} &= \frac{{\cal A}_{D-2}}{8\pi G} \int d\lambda R^{D-2}
\left [ \Delta K_0 +\frac{2 (D-2) \tilde{\alpha} }{(D-4)} \left( 2p_0 \Delta K_0 - 2 p_1 \Delta K_1
- \Delta (K_0K_1^2) \right) \right]\\
&= \frac{{\cal A}_{D-2}}{8\pi G} \int d\lambda R^{D-2} \Delta \left [ K_0
\left ( 1 + \frac{2(D-2)\tilde{\alpha}(1-f)}{(D-4)R^2} \right) - \frac{4(D-2)\tilde{\alpha} }{(D-4)} \frac{\ddot{R}}{R} K_1
\right].
\eeal$}
\ee
Noting that 
\be
\frac{f'\dot{\tau}}{2} = \dot{\tau} \left ( f\dot{\tau} K_0 + \ddot{R} \right) = 
\left ( 1 - \frac{\dot{R}^2}{f} \right) K_0 + \dot{\tau} \ddot{R},
\ee
we find that the combined bulk-term and wall integrals give
\be
\beal
I_{wall} = &\frac{{\cal A}_{D-2}}{8\pi G} \int d\lambda R^{D-2} \Delta \Biggl [
-\frac{4(D-2)\tilde{\alpha}}{(D-4)} \frac{\ddot{R}}{R} K_1\\
& \hskip 1cm + \left ( \frac{\dot{R}^2}{f} K_0 - \dot{\tau} \ddot{R}\right)
\left ( 1 + \frac{2(D-2)\tilde{\alpha}(1-f)}{(D-4)R^2} \right) \Biggr ].
\eeal\ee
Putting together, we see that the action of the bubble geometry is
\be
I_{bubble} = I_{wall} + I_C - S_{BH}(M_-)
\ee
hence the action for the tunnelling instanton is
\be\label{bubbleaction}
I_{\cal B} = I_{bubble} - I_{seed}
= S_{BH}(M_+) - S_{BH}(M_-) + I_{wall}.
\ee

Therefore, we have shown that the amplitude for vacuum decay including the 
higher derivative GB term in arbitrary dimensions has the same form
as the decay rate computed for vacuum decay in Einstein gravity in 4D. While
we leave the numerical evaluation of this action for arbitrary $M_+$ and $M_-$ 
for future study, we note that the contribution from the wall integral vanishes for
static instantons, which are the relevant instantons for Higgs vacuum decay \cite{Burda:2015isa,Burda:2016mou}.

The central result of this section is therefore that, as with Einstein gravity, 
the probability of thermal decay seeded by a black hole in EGB gravity is simply the difference in 
entropy of the seed and remnant black holes:
\be\label{probability}
{\cal P} \sim \text{exp} \left [ - (S_{seed} - S_{remnant} )/\hbar \right]
\ee

\section{The Hawking-Moss Instanton}\label{sec:HM instanton}

In the previous section, we derived the bounce action as the difference in entropies between the 
cosmological and event horizons. However, bubble nucleation is not the only decay mechanism for 
false vacuum decay.
Another process is the HM transition, that describes the field fluctuating from the false vacuum to the top of potential \cite{Hawking:1981fz}.
At the top of the potential, the field is at an unstable point and can roll down to either the false 
or true vacuum. The HM bounce is responsible for mediating the system's evolution into a new state, 
as illustrated in Fig.\ \ref{fig: potential}.
The HM transition depends only on the values of the potential energy densities at the top of the 
barrier and at the false vacuum, rather than the details of the potential in between. In 
\cite{Gregory:2020cvy,Gregory:2020hia}, the impact of black holes on the HM transition was studied, with the general 
finding that black holes enhanced the transition probability. In addition however, it was found that the
action could potentially become negative for very large seed black holes, unless a thermodynamically
motivation condition was applied -- \textit{Cosmological Area Conjecture}: that the 
cosmological horizon area can never increase in an up-tunnelling transition (see \cite{Gregory:2020hia} 
for a discussion and motivation for this principle).

The general Black-Hole-Hawking-Moss (BHHM) transition will have a `seed' black hole in the 
false vacuum and transition to a `remnant' black hole at the top of the potential. We change 
the notation in \eqref{Emetric} and \eqref{metricfunctions} of ``$f_\pm$'' to ``$f_{F/T}$'' 
to correspond to the labellings of the false vacuum ($F$) and the top of the potential ($T$). 
These will be characterised by mass parameters $\mu_{F/T}$, and vacuum parameters $\ell_{F/T}$, 
where $\ell^2=(D-1)(D-2)/2\Lambda$. 
The tunnelling rate is dominated by the Boltzmann factor:
\be
\Gamma_{FV \rightarrow Top}\sim e^{-B}.
\ee
From \eqref{probability}, we see that the change in action is the difference in the entropies, 
justifying the interpretation of the HM probability as a Boltzmann suppression:
\be
B_{FV \rightarrow Top}= I_T-I_F=[S_{CH}+S_{BH}]_F-[S_{CH}+S_{BH}]_T.
\label{HMaction}
\ee
Thus, the form of the BHHM action is the same as for Einstein gravity, however the entropies in 
the expression are now determined by the EGB formula:
\be\label{entropy}
S_i=\frac{ {\cal A}_{D-2}}{4 G} r_i^{D-2}
\left ( 1 + \frac{2(D-2) \tilde{\alpha}}{(D-4)r_i^2} \right)
\ee
where the subscript $i$ corresponds to the black hole or cosmological event horizon radius, 
which in general is also modified by the GB term. 

The likelihood of the BHHM instanton is obviously dependent on the details of the seed black hole 
and the dimension of the spacetime, but some general trends can be noted. For example, when a 
black hole is introduced, the total entropy drops (though note the exception we will discuss 
for 5D presently), thus for a given seed mass the BHHM transition will always prefer to jump 
to a pure de Sitter configuration, as this will maximise the negative contribution to \eqref{HMaction}. 
Similarly, introducing a seed black hole will also lower the action, since the overall entropy 
in the false vacuum will drop. Thus, just as with Einstein gravity, we expect that introducing 
seed black holes will also catalyse the HM transition.

\begin{figure}[ht]
\centering
\includegraphics[width=13cm]{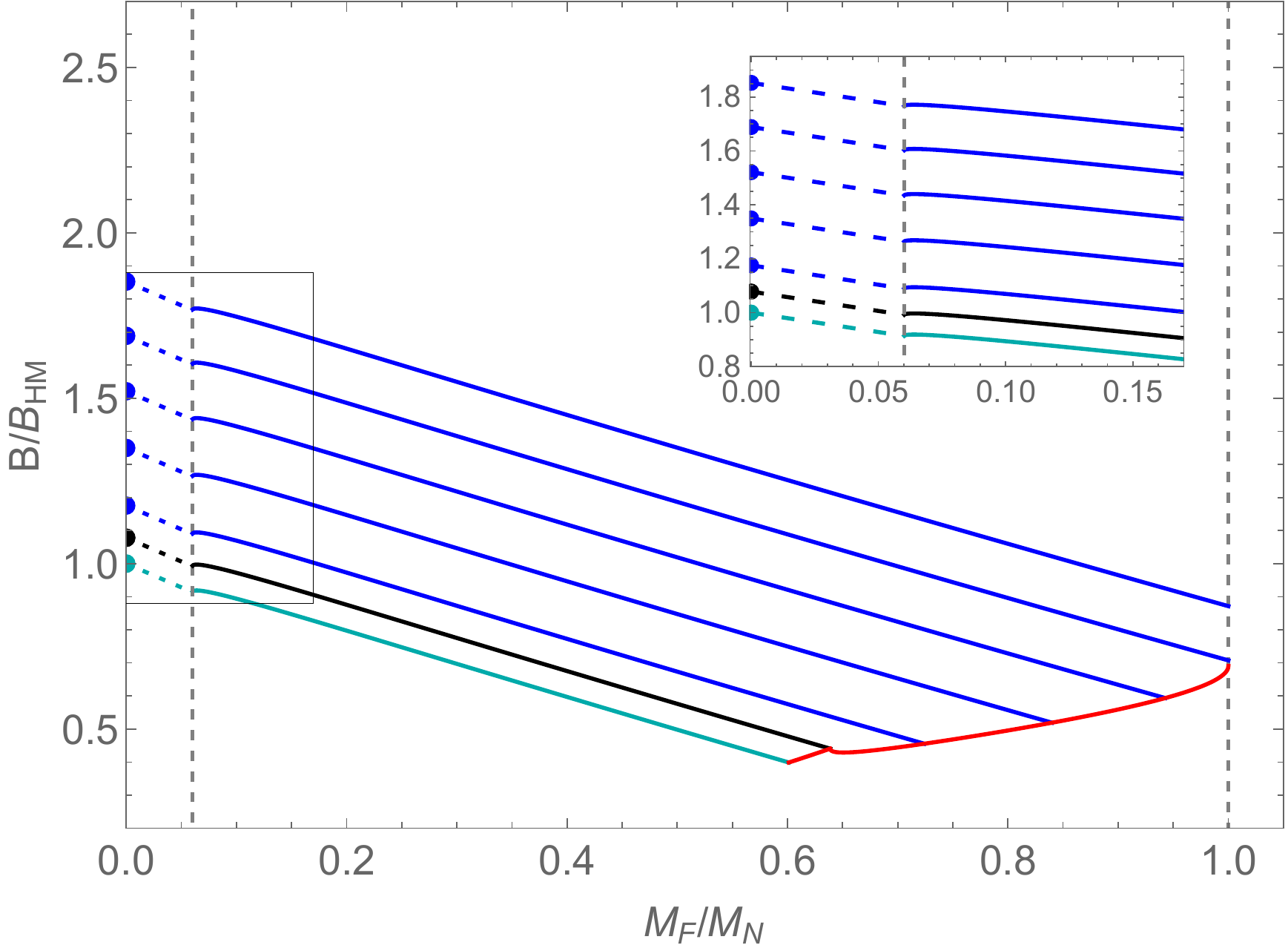}
\caption{The ratios of tunnelling components $B/B_{HM}$ change with the seed black hole mass in 5D spacetime, 
$M_F/M_N$, for fixed values of $\ell_F=5$, $\ell_T=4.5$, and $\alpha=0.2$ where the relative 
mass gap is $M_C\approx0.06M_N$. The blue solid lines represent remnant black hole masses ranging from 
lower to higher as ``$0.2 M_{N}$, $0.4 M_{N}$, $0.6 M_{N}$, $0.8 M_{N}$, $M_{N}$", respectively; 
while the black solid line represents a remnant critical mass $M_{C}$. The vertical dashed lines 
indicate the lower and upper bounds of the seed black hole mass. The red curve corresponds to 
the cosmological equal area limit. The cyan line represents a transition to a final de Sitter state 
without a remnant black hole, while the dots on the left axis show the transitions from pure de Sitter 
to remnant masses. The dashed lines on the left 
part of the graph, and the inset, represent a seed mass below the mass gap.}
\label{fig:HMD5}
\end{figure}

\begin{figure}[ht]
\centering
\includegraphics[width=13cm]{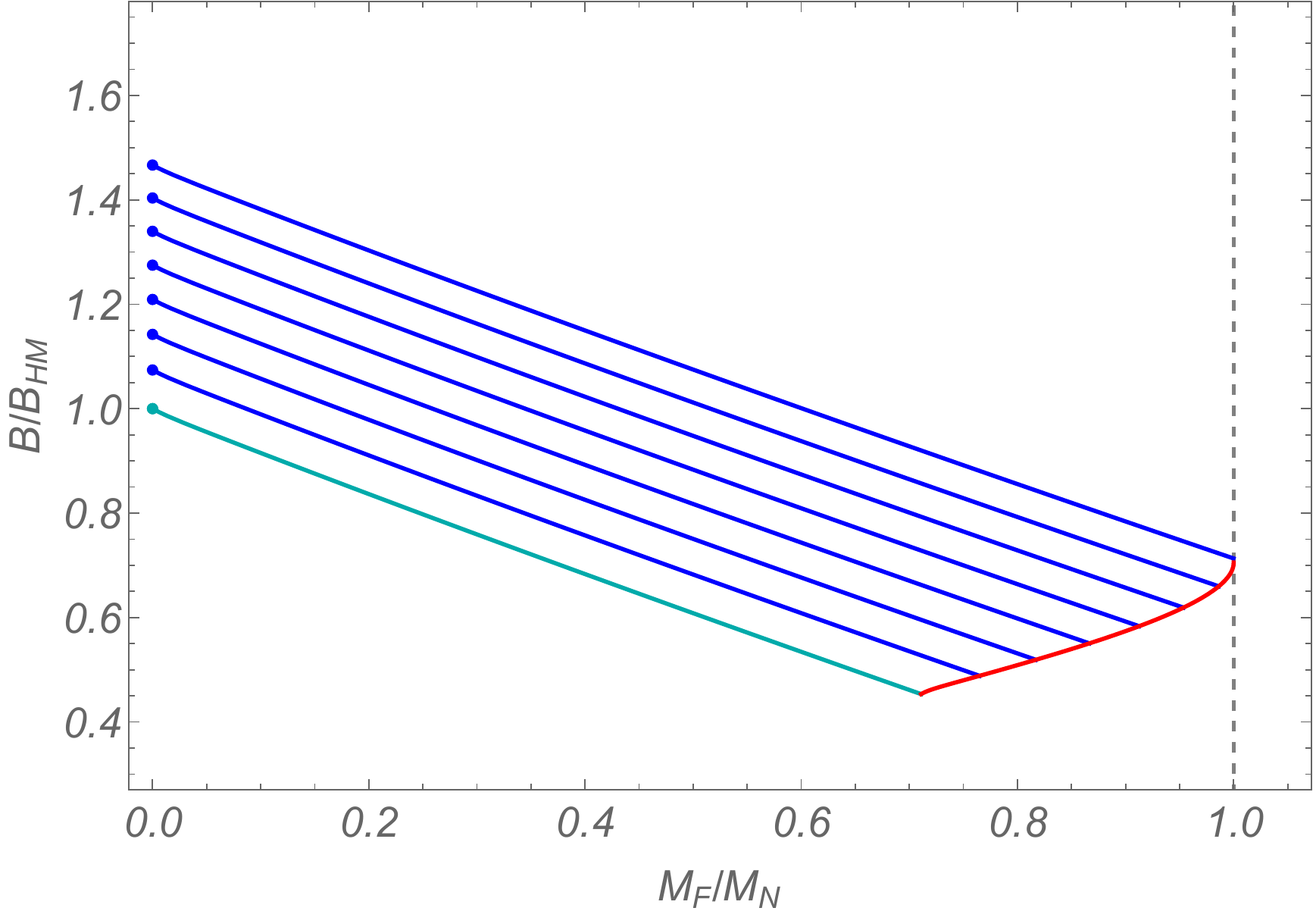}
\caption{The ratios of tunnelling exponents $B/B_{HM}$ as a function of seed mass in 6D spacetime, $M_F/M_N$, 
shown for parameter values $\ell_F=5,\ell_T=4.5$ and $\alpha=0.2$. The dashed vertical line marks 
the Nariai limit for the seed mass.  Blue solid lines represent different remnant black hole masses, 
ranging from $0.1 M_{N}$ to $0.7 M_{N}$ with steps of $0.1 M_{N}$ from lower to higher. The red curve gives the equal 
horizon area constraint. The cyan line represents a transition from a seed black hole to pure de Sitter spacetime. 
The dots represent a transition from pure de Sitter spacetime to a spacetime with a remnant black hole.}
\label{fig:HMD6}
\end{figure}

There is a limit however to how far we can drop the action, and this is determined by the \emph{Cosmological 
Area Principle} \cite{Gregory:2020hia}, that the area of the cosmological horizon must never increase:
\be
S_c |_T\leq S_c |_F 
\ee
As we saturate this bound, the HM action becomes the difference in entropy of the seed and remnant black holes.
Thus, we get broadly the same picture as for the Einstein case, where the preferred instanton 
(the one with lowest action) is a jump to a pure vacuum solution at the higher vacuum energy 
(i.e.~no remnant black hole) for lower seed masses. As the seed mass increases, we hit the 
Cosmological Area Bound and the transition is to a black hole geometry with the ``top'' vacuum 
energy (see figures \ref{fig:HMD5}, \ref{fig:HMD6}).

Therefore, broadly speaking the picture for the BHHM decay process looks qualitatively very similar 
for the EGB case as for the pure Einstein case across arbitrary dimensions with one interesting 
exception. Solving $f(r_h)=0$ for the black hole horizon 
at $r_h$ yields the following expression for the mass:
\be\label{massind}
M= \frac{(D-2) \mathcal{A}_{D-2}}{16\pi G} \mu
=\frac{(D-2) \mathcal{A}_{D-2} r_h^{D-5}}{16\pi G}
\left(\tilde{\alpha} + r_h^2-\frac{r_h^4}{\ell^2}\right).
\ee
For $D=5$, we see that while smaller black hole horizons correspond to lower black hole masses, 
it is not possible for $M\to 0$ while maintaining a horizon. Indeed,
inspecting the expression for $f(r)$ at the origin for $D=5$ 
yields $f(0) = 1 - \sqrt{\mu/\tilde{\alpha}}$, which 
becomes positive as $\mu$, hence $M$, becomes sufficiently
small: $GM<3\pi\alpha/4$, 
indicating the loss of an horizon. 
This is consistent with numerical studies of collapse \cite{Deppe:2012wk, PhysRevLett.114.071102}, which find that black holes will not form dynamically if the total mass-energy content of the spacetime is below a critical value in both asymptotically flat and asymptotically AdS spacetime.
This \emph{mass gap} was also
noted in \cite{Frolov:2015bta} in an exploration of black hole formation during gravitational collapse, and differences in thermodynamic phenomenology and gravitational collapse with GB corrections between five and higher dimensions were noted in 
\cite{Cai:2001dz, Wu:2021zyl}. Thus, the picture for the HM transition will be distinctive in 5 dimensions. We, therefore, present results for $D=5$ separately and present the HM 
results for $D=6$ as representative of higher dimensions.

\medskip

\noindent $\bullet$ 5D HM

Taking $f(r_h,r_c)=0$ and $D=5$ in \eqref{metricfunc} (noting $\tilde{\alpha} = 2\alpha$ in 5D) the black hole and cosmological event horizons can be obtained in a simple form:
\be\resizebox{0.90\hsize}{!}{$
r_h=\left(\frac{\ell^2}{2}-\sqrt{2 \alpha \ell^2+\frac{\ell^4}{4}-\frac{8GM \ell^2}{3\pi}}\right)^{\frac{1}{2}}, ~~~
r_c=\left(\frac{\ell^2}{2}+\sqrt{2\alpha \ell^2+\frac{\ell^4}{4}-\frac{8GM \ell^2}{3\pi}}\right)^{\frac{1}{2}}$}
\ee
Recall that $M_C = 3\pi \alpha/4$, is the critical minimum mass that allows for an event horizon to exist, and the Nariai mass, $M_N$, where the black hole and cosmological horizons merge  \cite{nariai1950some,Nariai1951OnAN} gives a maximum:
\be
GM_N=\frac{3\pi }{32} \left ( \ell^2 + 8\alpha\right).
\ee
Any solution with a mass parameter larger than $M_N$ will not
have the correct signature, hence is considered unphysical, thus all EGB 
\emph{black hole} solutions in $5D$ must satisfy $M_C \leq M \leq M_N$. 
Solutions with 
$M<M_C$ have no event horizon, but an examination of the geometry near $r=0$ shows that there is a solid angular deficit at the origin, which can be regularised by a similar method to the conical deficit regularisation, which allows one to integrate the Euclidean action, resulting in no additional contribution. The space of massive particle solutions in 5D EGB can therefore be thought of as 
analogous to AdS solutions in 3D Einstein gravity, where there is a finite range of masses 
for which the particle is a conical deficit, above which the solution becomes a BTZ black hole \cite{Witten:1988hc,Banados:1992wn}.

We depict the constraints on the black hole mass parameter in Fig.~\ref{fig:HMD5} by plotting 
the ratio of the BHHM instanton action to the pure HM instanton action, $B/B_{HM}$, as a function 
of the seed black hole mass, $M_F/M_{N}$. Here, $B_{HM}$ denotes the pure HM instanton without 
seed or remnant black holes, which only depends on the scale parameters of the initial and final states. The blue solid lines represent different remnant black hole masses, ranging from $0.2 M_{N}$ to $M_{N}$ in increments of $0.2 M_{N}$ from lower to higher, while the cyan line on the bottom represents a transition with no remnant black hole. The black line shows the lowest mass when a black hole horizon can exist.
As with the Einstein case, the BHHM action increases with the remnant mass, indicating larger 
remnant black holes slow down the transition. Two vertical dashed lines denote the lower and 
upper bounds of the seed mass for a black hole, which are constrained by mass gap and Nariai limits respectively. 
The red curve shows the limiting constraint of the cosmological area conjecture. In all cases, 
the Nariai limit is an upper bound on the seed mass. 

On the left hand side of Fig.~\ref{fig:HMD5}, the dashed lines indicate the region 
where the seed mass is smaller than $M_C$, hence there is no black hole horizon. There is still a 
mass at the origin however, and the cosmological horizon area is reduced as a result of the seed 
mass parameter, and linearly approaches the pure de Sitter area as $M\to 0$. Note that as $M$ increases 
from $M_C$, there is a small uptick in the action, reflected in the detail of the inset. This is 
due to the dependence of the entropies on $\Delta M = M-M_C$. The black hole horizon entropy 
$S_{BH} \propto \sqrt{\Delta M}$, whereas $S_{CH} \sim S_{DS} - {\cal O}(\Delta M)$, thus the 
overall entropy of the seed actually (temporarily) \emph{increases} as a result of adding mass 
to the system. Meanwhile, at the lower part of the plot along the red equal area line, the same 
phenomenon occurs, only this time causing a small downtick as the action of the remnant increases 
as the mass creeps above $M_C$.

\medskip

\noindent $\bullet$ 6D HM

Following the same procedure, Fig.~\ref{fig:HMD6} displays the ratio $B/B_{HM}$ as a function 
of seed black hole mass, $M_F$, for different remnant black hole masses, $M_T$, in $6D$. 
Unlike in $5D$ there is no mass gap, and only the Nariai limit sets an upper bound on the size 
of black holes. As before, the cosmological area conjecture gives a cut-off on the allowed parameter space. 

In Fig.~\ref{fig:HMD6}, we see a plot qualitatively the same as for the Einstein BHHM transitions: 
the action decreases as the seed mass increases, with the jump to a pure de Sitter spacetime at the top of 
the potential shown in cyan. The cosmological area conjecture provides a cutoff at larger seed 
masses, shown in red, where the transition now leaves a remnant black hole. The generic BHHM transition with both seed and remnant is shown in blue, where the remnant mass along a blue line is constant. The solid dots at the left depict the transition from pure de Sitter spacetime to a spacetime with a remnant black hole, especially, the cyan dot is the original HM transition.

\section{Four dimensions}\label{sec:4d}

Having derived results for vacuum decay with higher derivative gravity in general dimensions,
we now restrict to $D=4$ to explore whether there is an impact from the modification of
the action -- we have already confirmed that the bubble solutions are identical to 
Einstein gravity as expected in Sec.\ \ref{sec:bubble nucleated decay}, therefore it only 
remains to compute the effect of the GB term on the action.

\subsection{Bubble nucleated vacuum transitions}

Recall that the general instanton is (Euclidean) time dependent and has both seed and remnant 
black holes with action given by \eqref{bubbleaction} and the equation of motion:
\be
\beal
\dot{R}^2 &= 1 - \left ( \bar{\sigma}^2 + \frac{\bar{\Lambda}}{3} 
+ \frac{(\Delta\Lambda)^2}{144 \bar{\sigma}^2}\right) R^2 - \frac{2G}{R}
\left ( \bar{M} + \frac{\Delta M \Delta \Lambda}{24\bar{\sigma}^2} \right)
- \frac{(G\Delta M)^2}{4 \bar{\sigma}^2 R^4}\\
&= f_\pm - \left ( \omega_\pm R + \frac{\Delta M}{2\bar{\sigma}R^2}
\right)^2 = F(R)
\eeal
\label{walleom}
\ee
where
\be
\omega_\pm = \frac{\Delta \Lambda}{12\bar{\sigma}} \mp \bar{\sigma}
\label{omegapm}
\ee
and $f_\pm = 1 - 2M_\pm/r - \Lambda_\pm r^2/3$ is the Schwarzschild potential.

First, note that it is not possible to nucleate a bubble with a remnant black hole unless there is a seed. 
To see this, note that \eqref{walleom} implies that 
\be
f_+ \dot{\tau}_+ = \omega_+ R + \frac{\Delta M}{2\bar{\sigma}R^2} \geq 0
\ee
If we have no black hole seed, then $f_+ = 1- \Lambda_+ r^2/3$, and $\Delta M = -M_-$. The equation of 
motion, $\dot{R}^2 = F(R)$, must have two zeros of $F(R)$ (unless the solution is static, $F(R) \equiv 0$). 
However, 
\be\resizebox{0.90\hsize}{!}{$
F'(R) = f_+' - 2 \left ( \omega_+ R + \frac{\Delta M}{2\bar{\sigma}R^2} \right)
\left ( \omega_+ - \frac{\Delta M}{\bar{\sigma}R^3} \right)
= - \frac{2\Lambda R}{3} - 2 (f_+ \dot{\tau}_+)
\left ( \omega_+ + \frac{M_-}{\bar{\sigma}R^3} \right)
<0$}
\ee
hence $F(R)$ is a monotonically decreasing function of $R$ and cannot have two zeros.
We therefore conclude that vacuum decay by bubble nucleation either is a CDL bubble, or is seeded 
decay with an initial seed black hole, which will nucleate a bubble that may, or may not, have a 
remnant black hole.

Turning to the computation of the action for the bubble, there are two parts: the difference in 
black hole entropies, and the contribution from the bubble wall, which is not obviously vanishing.
Dealing with this latter term, the $\alpha$ dependent part of the contribution is $I_{wall,\alpha} 
= I^+_{{\cal W},\alpha} - I^-_{{\cal W},\alpha}$, where
\be
I^\pm_{{\cal W},\alpha} =  \frac{2\alpha}{G} \int d\lambda \left [ 
\left ( K_0^\pm- \frac{f'_\pm \dot{\tau}_\pm}{2}\right)
\left ( 1-f \right) - 2 R \ddot{R} K_1^\pm \right]
\ee
Here, $K_{0,1}$ were defined in \eqref{kays}, and for the wall trajectory \eqref{walleom} are
\be
K_0 ^{(\pm)} = \omega_\pm - \frac{\Delta M}{\bar{\sigma}R^3} \quad,\qquad
K_1 ^{(\pm)} = \omega_\pm + \frac{\Delta M}{2\bar{\sigma}R^3}
\ee
we therefore see that the $\alpha$ contribution to the wall action is:
\be
\resizebox{0.90\hsize}{!}{$\beal\label{int_w}
I^\pm_{{\cal W},\alpha} &=  \frac{2\alpha}{G} \int d\lambda \left [ 
\left ( K_0^\pm- \frac{f'_\pm \dot{\tau}_\pm}{2}\right)
\left ( 1-f \right) - 2 R \ddot{R} K_1^\pm \right]\\
&=\frac{2\alpha}{G} \int d\lambda \left [ 
\frac{K_0^\pm \dot{R}^2- R\ddot{R} K_1^\pm}{f_\pm} - K_0^\pm \dot{R}^2 
- R \ddot{R} K_1^\pm  \right]\\
&=\frac{2\alpha}{G} \int d\lambda \left [
\frac{\dot{R}^2}{f_\pm} \frac{d~}{d\lambda} \left ( \frac{\omega_\pm R}{\dot{R}}
+ \frac{\Delta M}{2\bar{\sigma}\dot{R} R^2} \right)
- \omega_\pm  \frac{d(R\dot{R}) }{d\lambda} +
\frac{\Delta M}{2\bar{\sigma}R^3} (2\dot{R}^2 - R \ddot{R})\right]
\eeal$}
\ee
But now we note that the middle term vanishes due to the periodicity of $R$,
the final term is the same on each side of the wall which will therefore cancel. Turning to the first term, this
can be rewritten using the equation of motion for the wall:
\be
\frac{\omega_\pm R}{\dot{R}}
+ \frac{\Delta M}{2\bar{\sigma}\dot{R} R^2}
= \frac{\sqrt{f_\pm - \dot{R}^2}}{\dot{R}}=
\text{sign($\dot{R}$)}\,
\sqrt{\frac{f_\pm}{\dot{R}^2}-1}
\ee
thus the first part of the integral 
in \eqref{int_w} becomes:
\be
\int \text{sign($\dot{R}$)}\,
\frac{\dot{R}^2}{f_\pm} \frac{d~}{d\lambda} \left ( 
\sqrt{\frac{f_\pm}{\dot{R}^2}-1} \right)\,d\lambda 
= \int \text{sign($\dot{R}$)}\,
\text{arctan} \left ( 
\sqrt{\frac{f_\pm}{\dot{R}^2}-1} \right)\,d\lambda 
\ee
due to the periodicity of the solution $R(\lambda)$, the arctan function is symmetric 
around the turning points of the solution, whereas $\dot{R}$ changes sign, thus this integral
vanishes and there is no $\alpha$ dependent contribution from the wall.

Examining the contribution of the entropies to the action reveals the presence
of an $\alpha-$dependent, but constant, contribution from each black hole horizon in 
\eqref{bubbleaction}. The entropy term, $S_{BH}$ is
\be
S_{BH} = \frac{\pi r_+^2}{G} \left ( 1 + \frac{4\alpha}{r_+^2} \right).
\ee
From the perspective of black hole thermodynamics, this constant shift is irrelevant 
\cite{Clunan:2004tb}, but for our action, it is potentially important. If we have both a seed 
and remnant black hole, these constant terms will cancel, but if we have a seed black hole 
that is wiped out by bubble nucleation, then there will be a residual $\alpha$ term in action.

We therefore arrive at the result that the bubble action in 4D is identical to the Einstein
action if there is either {\it both} a seed and remnant black hole or {\it neither} a seed
nor remnant black hole. However, if we have a seed black hole and no remnant, then
the GB term leaves an imprint on the action:
\be 
I_{{\cal B},NR} = \frac{\pi r_+^2}{G} + \frac{4\pi\alpha}{G} +I_{wall} 
\ee

Thus, for bubble nucleated decay, the GB term either has no impact or actually
suppresses tunnelling ($I_{{\cal B}}$ is increased), in the case that the decay
removes the black hole altogether. This is perhaps not surprising, once one considers the 
fact that the GB term is a topological invariant in 4D, thus it should only impact on topology 
changing transitions. Perhaps more surprising is that the GB term inhibits topology changing decay.

\subsection{Hawking-Moss Tunnelling}

Armed with the results of the previous subsection, we can now easily deduce how
the GB term impacts 4D HM transitions. If there is no topology change, there is no change in the 
action: i.e.\ for the pure HM instanton and for a transition with a seed black hole to a remnant 
black hole geometry, which corresponds to the (red) Cosmological Area Principle boundary. However, 
for the HM transitions that jump from a Schwarzschild de Sitter (SdS) spacetime to a pure de Sitter spacetime, there is an 
$\alpha-$dependent contribution that suppresses this topology changing transition. We have the 
bizarre situation that the universe preferentially jumps to an SdS solution with a vanishingly 
small mass -- this will have the same topology as the initial SdS state, but the tiny black hole 
will presumably instantaneously evaporate leaving effectively pure de Sitter at the top of the potential. 
Whether or not such a Planckian sized black hole should be included in a semi-classical description 
is a very good question! Finally, unlike the bubble nucleation, it is possible for a HM transition 
to occur from pure de Sitter false vacuum to an SdS spacetime at the top of the potential. For these 
(topology changing) configurations the action is now lowered by $\delta I \sim 4\alpha/G - 
2\pi \ell M$ relative to the pure HM transition, thus again there is a preferred transition to 
a vanishingly small mass black hole universe.

\section{Discussion}\label{sec:discussion}

In this paper, we studied the impact of higher order terms in the gravitational action, focussing 
on the Gauss-Bonnet invariant to maintain well-posedness of the equations of motion. We considered 
both first order, tunnelling, vacuum decay transitions as well as Hawking-Moss jumps. For bubble 
nucleation, we derived the general equations of motion for a bubble, including the GB term in 
arbitrary dimension, and found that the general result of \cite{Gregory:2013hja}, that larger black holes catalysed 
a static instanton with the action determined by the entropy difference between the seed and remnant 
black hole remains true with the GB term, and in all dimensions. The HM transitions were 
algebraically easier to explore, with results qualitatively similar to the Einstein process.

Note that our results have been obtained within the 
\emph{thin wall} approximation, in which the vacuum transitions completely and instantaneously 
across the wall. While we do not expect higher derivative terms to significantly change the discussion for scalar field decay -- in particular thermal corrections -- in 4D, the impact of the changed geometry in higher dimensions would be interesting to investigate.

In 4D, the fact that the GB term is a topological invariant means that it will not impact on instantons 
that have both seed and remnant black holes, however, when there is a transition from a seed black hole 
with no remnant, the GB term suppresses the transition (for positive $\alpha$). For a situation with no seed black hole, only 
the HM transition can result in a geometry with a remnant black hole, and here the GB term enhances the 
decay, although the lack of continuity of the action as a function of black hole mass means that an 
arbitrarily small mass black hole would have the lowest action, likely taking the spacetime outside 
of the semi-classical regime.
The special case of 5D, with its mass gap for $\alpha>0$, led to some interesting additional features in the transition amplitudes, and this case may merit further examination.

Interestingly, there are problems with a generic higher order curvature terms in the action. Without 
the well-posedness of these Lovelock terms, the higher order derivatives mean that singular instantons 
cannot be regularised in a well-defined and rigorous manner without a UV completion. It is also likely 
the case that the generalised Birkhoff theorems which specify the form of bubble transitions are no longer 
applicable in the presence of such terms \cite{Oliva:2011xu,Oliva:2012zs}.

Finally, it would be interesting to return to the analysis of pure EGB vacua of \cite{Charmousis:2008ce}, and to consider more general black hole bubble solutions. The cubic Friedmann equation \eqref{cubicFriedmann} would need to be solved (with appropriate modifications for including the GB branch), and without the link to the Einstein limit of the Einstein branch, all roots of the cubic might be equally valid. 

What we have learned is that Lovelock terms appear to give a very similar picture to the black hole 
catalysed decay as the Einstein case, albeit with greatly more convoluted algebra!

\acknowledgments

We would like to thank Christos Charmousis and Sam Patrick for discussions and useful suggestions.
This work was supported in part by the STFC Consolidated Grant ST/P000371/1 (RG), the Chinese Scholarship Council and King's College London (SH), the Aspen Center for Physics, supported by National Science Foundation grant PHY-2210452 (RG),
and the Perimeter Institute for Theoretical Physics (RG). 
Research at Perimeter Institute is supported by 
the Government of Canada through the Department of Innovation, 
Science and Economic Development Canada and by the Province of 
Ontario through the Ministry of Colleges and Universities.

\bibliographystyle{JHEP.bst}
\bibliography{refs}

\end{document}